\documentclass[11pt,twoside,a4paper]{article}

\usepackage{authblk}
\usepackage{graphicx}
\usepackage{amsmath}
\usepackage{amssymb}
\usepackage{mathrsfs}
\usepackage{color}
\usepackage{hyperref}

\newcommand{\ie}{\textit{i.e.}~}
\newcommand{\etal}{\textit{et al.}}

\definecolor{blue}{rgb}{0,0,1}
\definecolor{green}{rgb}{0,0.5,0}
\definecolor{red}{rgb}{1,0,0}
\definecolor{pink}{rgb}{0.9,0.3,0.7}
\definecolor{azur}{rgb}{0,0.5,0.5}
\definecolor{orange}{rgb}{1,0.5,0.2}
\definecolor{brown}{rgb}{0.5,0,0}

\begin{document}

\title{Graph matching based on similarities in structure and attributes}
\author[1]{Raphaël Candelier}
\affil[1]{Sorbonne Université, CNRS, Institut de Biologie Paris-Seine (IBPS), Laboratoire Jean Perrin (LJP), F-75005, Paris}
\date{September 30, 2024}

\maketitle


\begin{abstract}

\noindent Finding vertex-to-vertex correspondences in real-world graphs is a challenging task with applications in a wide variety of domains. Structural matching based on graphs connectivities has attracted considerable attention, while the integration of all the other information stemming from vertices and edges attributes has been mostly left aside. Here we present the Graph Attributes and Structure Matching (GASM) algorithm, which provides high-quality solutions by integrating all the available information in a unified framework. Parameters quantifying the reliability of the attributes can tune how much the solutions should rely on the structure or on the attributes. We further show that even without attributes GASM consistently finds as-good-as or better solutions than state-of-the-art algorithms, with similar processing times.

\end{abstract}


\section{Introduction}

The importance of graph matching comes from the fact that a considerable amount of phenomena with very diverse nature can be represented with the same concept of graph (or networks), and the ability to find correspondences between their atomic elements -- vertices and edges -- has concrete application in a number of domains including computational biology \cite{Junker_2007,Emmert_2011,Devkota_2024}, neuroscience \cite{Olivetti_2016,Mheich_2020,Pedigo_2023}, chemoinformatics \cite{Randic_1979,Willet_2011,Lin_2022}, medical imaging \cite{Oyarzun_2022}, computer vision \cite{Conte_2003,Hsieh_2008,Sun_2020}, machine learning \cite{Li_2019,Yan_2020} and linguistics \cite{Mehler_2010}.

As the problem of matching graphs structures is NP-complete \cite{Bunke_2000}, finding the optimal solution would require exponential time and space and an abundant litterature aims at finding in a reasonable time either isomorphic relations for exact graph matching or approximated solutions for matching graphs in an error-tolerant way (see \cite{Bunke_2000,Conte_2011,Emmert_2016} for reviews). Actually the graph matching problem can be recast as a special case of \textit{quadratic assignment problem} (QAP) \cite{Umeyama_1988}, like the famous traveling salesman problem and many other problems in combinatorial optimization and distributed resource allocation.

On the algorithmic side, 2opt -- a simple local optimizer that swaps pairs recursively -- is one of the oldest algorithm still in use today~\cite{Croes_1958}. Many approaches try to reduce the problem down to a \textit{linear assignment problem} (LAP) \cite{Blondel_2004,Zager_2008,Nikolic_2012}, for instance by defining explicitely a score matrix representing the similarity between graph elements and applying a LAP solver like the popular Jonker-Volgenant algorithm~\cite{Jonker_1987}, which operates in polynomial time \cite{Crouse_2016}. Vogelstein~\etal~\cite{Vogelstein2015} have proposed a very efficient algorithm for approximated solutions, \textit{Fast Approximate QAP} (FAQ), that first solves a relaxed, linearized version of the QAP and subsequently projects the solution back onto the permutation space. FAQ has then been implemented in Scipy and it is still considered today as a standard.

However, most real-world graphs have \textit{attributes} attached to their edges or vertices \cite{Emmert_2016}. For instance, in the context of protein-protein interaction networks the vertices are characterized by a unique protein identifier while the edges may bear multiple association weights for different quantifications of the interactions~\cite{Szklarczyk_2023}. Let us thus define attributes as functions that can account for virtually any property of vertices and edges, with values that can be numerical or not. Matching the graph structure (\ie just the connections backbone) can be done independantly of the attributes but the solutions are necessarily less accurate. Several existing algorithms can actually manage some attributes -- for instance FAQ accepts inputs with one weight per edge -- but still, almost all the algorithms developped so far can only use just a part of the available information of real-world graphs. This is detrimental from a strategic point of view as the information bear in attributes can greatly improve both the solutions and the searching time; for instance Dickinson \etal \cite{Dickinson_2004} proved that for graphs possessing unique vertex labels the computational complexity is only quadratic in the number of nodes.

Here, we propose a new algorithm termed \textit{Graph Attributes and Structure Matching} (GASM) that is precisely designed to integrate any amount and any type of attributes. It has been largely inspired by iterative methods and especially a series of three articles: Kleinberg's HITS algorithm \cite{Kleinberg_1999} that projects any graph on the so-called ``hubs and authorities'' graph, the generalization by Blondel \etal~\cite{Blondel_2004} that adapts the same idea to calculate iteratively vertex similarity scores and match any pair of graphs, and finally the work of Zager and Verghese \cite{Zager_2008} who elegantly introduced an edge score matrix to fix the convergence issues and the dependency on initial conditions. Zager \etal~also tried to incorporate the handling of a categorical vertex attribute, but the resulting algorithm is not robust.

In GASM all the contraints related to the attributes are introduced \textit{a priori}, which creates a coupling between the attributes and the structure during the iterations. Error parameters for each attribute allow to tune how much the solutions should rely on the structure or on the attributes. Interestingly, a small amount of noise is also introduced to lift the degeneracies due to local symmetries and further improve the general quality of the solutions. Finally, 
a simple convergence criterion is introduced to limit the number of iterations.

This paper is organized as follows. In Section \ref{sec:preamble} some general ideas, definitions and notations are introduced, while Section \ref{sec:algorithm} presents the GASM algorithm. Results are presented in Section \ref{sec:results}, with a comparison with Zager (\ref{sec:degeneracy_handling}) and benchmarks on isomorphic matching (\ref{sec:isomorphic_matching}), QAPLIB (\ref{sec:QAPLIB}), graph degradation (\ref{sec:degradation}) and speed (\ref{sec:speed}). The paper ﬁnishes in Section \ref{sec:conclusion} with a conclusion.


\section{Preamble: definitions, notations and general observations}
\label{sec:preamble}

Let us consider the comparison of two graphs $G_A$ and $G_B$, which may be both directed or both undirected but that are not multigraphs. We index variables and matrices of the corresponding graphs with $A$ and $B$, and with $\ast$ as a replacement symbol for quantities that are defined similarly in both graphs; for instance the number of vertices is $n_\ast$, meaning that it is $n_A$ for $G_A$ and $n_B$ for $G_B$. Let us also note the number of edges $m_\ast$, among which there are $\mu_\ast$ self-loops, and the adjacancy matrix $\Lambda_\ast$. Variables are also used without index when they are equal for both graphs: for instance we may refer to the number of vertices $n$ when $n = n_A = n_B$.

\subsection{Attributes}

The graphs considered here can have any number of vertex or edge attributes. Let us separate the \textit{measurable} attributes from the non-measurable, or \textit{categorical} ones. There are many cases where this distinction is obvious: in the case of neural networks for instance~\cite{Meyer-Baese_2014}, the activation functions belong to a set of functions of different families and can thus be considered as a categorical attribute of vertices, which means it can only be compared for exact correspondence or not. In contrast, the edges weights and vertices biases have values in $\mathbb{R}$ and a distance can be defined, so they are measurable attributes; indeed, a weight of 0.1 is strictly different from 0 and 1, but in a matching context it is natural to consider that is is better matched with the former.

However, an attribute with numerical values does not automatically belong to the measurable class, since it can represent indexes for instance. The measurability of any attribute is specific to the graphs of interest and has to be determined accordingly by the end user.

\subsection{Accuracy}

As one usually want to associate the nodes, if $n_A \ge n_B$ there are $\frac{n_A !}{(n_A-n_B) !}$ possible association sets, a potentially prodigious amount among which a few may be meaningful matchings while the vast majority is composed of non-sensical matchups. A key factor for finding good matchings and to benchmark algorithms is the ability to compare candidate solutions.

A common measure is the \textit{accuracy} of a matching $\mathcal{M}$, defined as the proportion of vertices pairs corresponding to the ground truth, and noted here $\gamma (\mathcal{M})$ or simply $\gamma$.

Importantly, the matching accuracy can go up to 1 for some pairs of isomorphic graphs, but not always, and when the graphs are non-isomorphic this is generally not true. In some cases the maximal possible average accuracy can be computed independantly of the matching algorithm (see for instance section \ref{sec:isomorphic_matching}), and for graphs with many local symmetries it can be arbitrarily low. So this is a delicate quantity to manipulate as in the general case the maximum possible average value is unknown and the values have no absolute meaning; one can only compare different accuracies relatively to each other, without knowing up to what point these solutions can be further improved. There are also very counterintuitive cases, like pairs of graphs for which any matching algorithm returns solutions with the same average accuracy -- see for instance the case of circular ladders in section \ref{sec:isomorphic_matching}.

\subsection{Matching quality}
\label{sec:matching quality}

But probably the most obvious limitation of accuracy is that is necessits to have the ground truth, which is by definition unknown in all real-world applications. Still, comparing the local properties of all the matched pairs to compute global quantities can always be done, and we shall refer to these global quantities as the matching \textit{qualities}. Several definitions of qualities can be derived, depending on the assessed local property.

Let us define the \textit{structural quality} $q_S$ as a measure of the local structural similarity between pairs of vertices. For a given matching $\mathcal{M}$, let $M$ be the binary matrix of size $n_A \times n_B$ such that:
\begin{align}
  [M]_{ij} & = \begin{cases}
    1, & \text{if node $i$ in $G_A$ is matched with node $j$ in $G_B$}\\
    0, & \text{otherwise}
  \end{cases}
\end{align}

Let us define:

\begin{equation}
  Z = \Lambda_A M - M \Lambda_B
  \label{eq:qs_Z}
\end{equation}

The structural quality $q_S$ of the matching then reads:

\begin{equation}
  q_S = \begin{cases}
    0 \qquad \textrm{if~} m_A = m_B = 0 \textrm{, otherwise:}\\[10pt]
    1 - \frac{tr \left( Z^\top Z \right)}{m_A + m_B} \qquad \textrm{for directed graphs}\\[10pt]
    1 - \frac{tr \left( Z^\top Z \right)}{2(m_A +m_B) - \mu_A - \mu_B} \qquad \textrm{for undirected graphs}
  \end{cases}
  \label{eq:strucural_quality}
\end{equation}

The idea behind this definitions is to count all the edge mismatches, defined as edges whose terminating vertices have matchups that are not themselves connected with an edge in the other graph. The intermediary matrix $Z$ has elements $z_{ij}$ set to $0$ when the pair of vertices $(i,j)$, respectively from $G_A$ and $G_B$, both have or both don't have a neighbor associated with the other vertex, and $\pm1$ otherwise. The trace of the product $Z^\top Z$ is a way to compute the grandsum of the squares of each element of $Z$, which is the number of discrepancies contained in $Z$. $q_S$ is then a scalar bounded in $[0,1]$, a higher value indicating a better overall matching of the local structure.

\begin{figure}[!b]
  \centering
  \includegraphics[width=1\columnwidth]{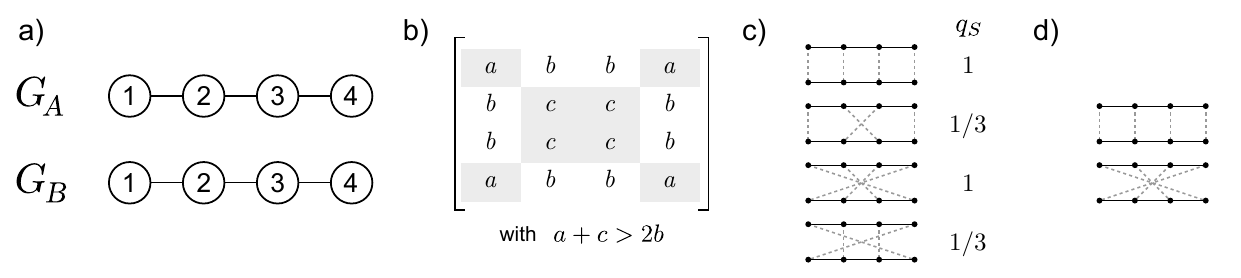}
  \caption{Exemple of matching degeneracy introduced by the score matrix.
  \textbf{a)} The graphs to match.
  \textbf{b)} Form of the score matrix returned by any algorithm exploiting the graph structure. The best matching solutions are composed of the grayed cells exclusively.
  \textbf{c)} The 4 matchings with a maximum total score of $2(a+c)$, along with their structural quality $q_S$.
  \textbf{d)} The only 2 matchings respecting structural correspondence.}
  \label{fig:ex_undet_0}
\end{figure}

Of course, it would be impractical to compute all the possible matchings and sort them by their structural quality $q_S$, so this cannot constitute the core of a brute force matching algorithm. Yet, it is an useful quantity as it allows to raise a certain type of degeneracy introduced by score matrices, which is a pivotal element for any LAP-based approach. A very simple illustration is given in fig.\ref{fig:ex_undet_0}, where a simple linear graph is matched with itself ($G_A = G_B$); let us call the vertices 1 and 4 of this exemple the \textit{side} vertices while vertices 2 and 3 are the \textit{inner} ones. Any algorithm exploiting the graph structure would give a score matrix with the form displayed in fig.\ref{fig:ex_undet_0}-b: as the side vertices are undistinguishable and the inner vertices as well, there can be only 3 different values in the score matrix: $a$ standing for the side-side scores, $b$ for side-inner scores and $c$ for inner-inner scores. In addition, $a+c$ should be greater than $2b$ if the local structural similarity is favored in the score determination. The point is that the score matrix cannot handle 4-point interactions and in this exemple there are 4 solutions with the same maximal scores $s=2(a+c)$ (fig.\ref{fig:ex_undet_0}-c), among which only 2 have a perfect structural quality $q_S$. So, without further processing the user has a $1/2$ probability to end up with a solution that is structurally unsound, because of the limitations of the score matrix and more generally that graph matching is treated as a LAP and not a QAP. We will see later in section \ref{sec:degeneracy_handling} how GASM is able to circumvent this limitation of LAP-based approaches.

Similarily, it is possible to define a matching quality for any attribute, but we will not use it in this paper so let us jump directly to the desription of the GASM algorithm.

\section{Algorithm}
\label{sec:algorithm}

\subsection{Attribute matrices}

We place ourselves in the case where the vertices and edges of $G_A$ and $G_B$ can have several attributes, some being measurable and some categorical. However, if one vertex (resp. edge) has a value for a given attribute all the other vertices (resp. edges) of both graphs should also have a value for this attribute.

Let us consider an attribute $\mathscr{A} \colon \theta \mapsto a(\theta)$ where $\theta$ can be a vertex or an edge without loss of generality. If $\mathscr{A}$ is categorical, the comparison of two elements $\theta$ in $G_A$ and $\theta'$ in $G_B$ can only have a binary outcome (similar or dissimilar) and is naturally represented by a boolean \textit{attribute distance matrix} $\mathcal{A}$ of size $n_A \times n_B$ (vertex attribute) or $m_A \times m_B$ (edge attribute) defined by:

\begin{align}
[\mathcal{A}]_{\theta\theta'} & = \delta_{a(\theta)a(\theta')} = \begin{cases}
  1, & \text{if}\ a(\theta) = a(\theta') \\
  0, & \text{otherwise}
\end{cases}
\label{eq:nm_attr_matrix_noerror}
\end{align}

\noindent where $\delta_{ij}$ is the Kronecker symbol.

However, in real-world applications there may be mistakes on some attribute values. For each attribute $\mathscr{A}$ let us introduce a positive scalar $\rho$ estimating the error over its values, a parameter that should be defined by the end user. If there is no noise or other source of uncertainty in the process used to obtain $a(\theta)$ and $a(\theta')$, \ie in the limit where $\rho \rightarrow 0$, $\mathcal{A}$ can be simply computed with eq.(\ref{eq:nm_attr_matrix_noerror}). In the limit where $\rho \rightarrow \infty$, there is no way to decipher which attribute values are correct and all the values in $\mathcal{A}$ should be equal and non-zero. So, in the general case let us consider the following definition:

\begin{align}
  [\mathcal{A}]_{\theta\theta'} & = \begin{cases}
    1, & \text{if}\ a(\theta) = a(\theta') \\
    e^{-\frac{1}{2\rho^2}}, & \text{otherwise}
  \end{cases}
\label{eq:nm_attr_matrix}
\end{align}

\noindent where the error parameter $\rho$ tunes the contrast in $\mathcal{A}$.

Now, if the attribute $\mathscr{A}$ is measurable the comparison between $a(\theta)$ and $a(\theta')$ can be a real scalar. In order to combine several distance matrices altogether it is preferable to keep them bounded in $[0,1]$, with 0 indicating dissemblance and 1 indicating similarity. Let us define the corresponding attribute distance matrix by:
\begin{align}
[\mathcal{A}]_{\theta\theta'} & = e^{\textstyle -\frac{[a(\theta)-a(\theta')]^2}{2\rho^2} }
\label{eq:m_attr_matrix}
\end{align}

In the limit where $\rho \rightarrow 0$, $\mathscr{A}$ become similar to a categorical attribute with as many categories as values and $\mathcal{A}$ can be computed using (\ref{eq:nm_attr_matrix_noerror}) as well. Otherwise, it is important to keep $\rho$ as close as possible of the real error to ensure the accuracy of the final solutions does not drop artificially. In case $\rho$ cannot be estimated a safe replacement is the standard deviation $\sigma_a$ of the distances $a(\theta)-a(\theta')$ for measurable attributes and of $\delta_{a(\theta)a(\theta')}$ for categorical ones, computed over all possible pairs $(\theta, \theta')$. $\sigma_a$ is indeed a higher bound for the estimation of the error.

Then, if there are $\zeta$ vertex attributes with associated distance matrices $(\mathcal{A}_1, \mathcal{A}_2, ..., \mathcal{A}_\zeta)$ and $\xi$ edge attributes with distance matrices $(\bar{\mathcal{A}}_1, \bar{\mathcal{A}}_2, ..., \bar{\mathcal{A}}_\xi)$, the global \textit{vertex distance matrix} $V$ and \textit{edge distance matrix} $E$ are defined as:
\begin{align}
V & = [\nu_{uv}] = J_n \odot \mathcal{A}_1 \odot \mathcal{A}_2 \odot ... \odot \mathcal{A}_\zeta \\
E & = [\epsilon_{ij}] = J_m \odot \bar{\mathcal{A}}_1 \odot \bar{\mathcal{A}}_2 \odot ... \odot \bar{\mathcal{A}}_\xi
\label{eq:attributes_prod}
\end{align}

\noindent where $J_n$ and $J_m$ are unit matrices of respective sizes $n_A \times n_B$ and $m_A \times m_B$, $\odot$ stands for the Hadamard product and $(u,v)$ and $(i,j)$ are pairs of vertices and egdes from $G_A$ and $G_B$. As all elements of all distance matrices stand in the interval $[0,1]$, all the $\nu_{uv}$ and $\epsilon_{ij}$ are also bounded in $[0,1]$.

\subsection{Scores}

The matrices $V$ and $E$ are akin to score matrices that can readily be used to match either the vertices or the edges, without taking into account any structural information on the graphs. For instance, if at least one of the graphs has no edge, a matching based on vertices attributes can be obtained by simply feeding $V$ in a LAP solver and searching for a maximum score matching.

However, in general it is desirable to take into account the similarities in both the structure and the attributes. Several algorithms have been designed to output a score matrix based on structural similarities, again to be supplied to a LAP solver. GASM integrates both information by using the vertex and edge distance matrices as initial conditions for an iterative procedure inspired by Zager \etal~\cite{Zager_2008}.

In the sequel we separate the cases of undirected and directed graphs. Since undirected graphs can easily be converted to directed graphs without loss of information it may seem sufficient to cover only the directed case. However, directed versions of undirected graphs are a peculiar subset of directed graphs for which a slightly different formalism can be applied and that has specific properties. We will see in the Results section that there are significant differences in matching accuracy and structural qualities, at least for all the algorithms considered here. In addition, for several algrithms including GASM, using directed versions of undirected graph requires twice the number of operations without any gain in return, so there is also a performance boost in separating the cases.

\subsection{Iterative procedure for undirected graphs}
\label{sec:iter_ug}

Let us first cover the case where both $G_A$ and $G_B$ are undirected graphs. For any vertice $v$, let $C_\ast(v)$ be the set of edges that are connected to $v$, and for any edge $i$ let $D_\ast(i)$ be the set of vertices it connects. $D_\ast(i)$ contains at most 2 vertex indices, and only one if $i$ is a self-loop. At iteration $k$, let $x_{uv}(k)$ denote the vertex similarity score between vertex $u$ in $G_A$ and vertex $v$ in $G_B$, and $y_{ij}(k)$ the edge similarity score between edge $i$ in $G_A$ and edge $j$ in $G_B$.

As an initial step, the vertex scores are defined as:

\begin{equation}
x_{uv}(1) = \left( \nu_{uv} + h_{uv}\right) \sum_{\substack{i \in C_A(u) \\ j \in C_B(v)}} \epsilon_{ij}
\label{eq:ug_init_dvlp}
\end{equation}

\noindent where $h_{uv}$ are random values drawn from the continuous uniform distribution between $0$ and a parameter $\eta \ll 1$. The role of this minute positive ``noise'' term is to help lifting degeneracies, and will be discussed later on in section~\ref{sec:degeneracy_symmetries}.
 
Then, for each iteration step $k>1$ the update equations are:

\begin{align}
y_{ij}(k) & = \frac{1}{f_y} \sum_{\substack{u \in D_A(i) \\ v \in D_B(j)}} x_{uv}(k-1)
\label{eq:ug_update_dvlp_Y}\\
x_{uv}(k) & = \frac{1}{f_x} \sum_{\substack{i \in C_A(u) \\ j \in C_B(v)}} y_{ij}(k)
\label{eq:ug_update_dvlp_X}
\end{align}

\noindent where $f_x$ and $f_y$ are normalization coefficients. These coefficients can be set to any positive finite value at any iteration without altering the outcome of the whole algorithm, since only the relative values of the scores are important for the LAP. To simplify the formulas we set $f_x=f_y=1$ in the sequel; however, it is worth noting that during numerical computation some normalization may be used to avoid floating point overflow. This is discussed in more details in section~\ref{sec:normalization}, where an approximated normalization factor is introduced for this practical purpose.

Let us now express these equations in a concise form using only elementwise matrix operations and matrix multiplication. Let the \textit{unoriented incidence matrix} $R_\ast$ be defined by:
\begin{align}
[R_\ast]_{ui} & = \begin{cases}
  1, & \text{if}\ i \in C_\ast(u) \\
  0, & \text{otherwise}
\end{cases}
\label{eq:def_u_incidence}
\end{align}

Each column of $R_\ast$ stands for an edge and has exactly two non-zero elements, except for self-loops which have a unique non-zero element. The initialization and update equations can then be written as:
\begin{align}
X_{1} & = (V + H) \odot (R_A E R_B^\top) \label{eq:ug_init_mat} \\[10pt]
Y_{k} & = R_A^\top X_{k-1} R_B \label{eq:ug_update_mat_Y_tmp} \\
X_{k} & = R_A Y_k R_B^\top \label{eq:ug_update_mat_X_tmp}
\end{align}

\noindent where $H$ is the matrix composed of the noise terms $h_{uv}$.

To reduce computation speed, two strategies can be employed: parallelization on GPU and using graphs complements on CPU. A GPU version of GASM has been implemented using eq.(\ref{eq:ug_init_dvlp}-\ref{eq:ug_update_dvlp_X}) and provide the best speed (see section~\ref{sec:speed}) for weakly connected graphs. A CPU version based on eq.(\ref{eq:ug_init_mat}-\ref{eq:ug_update_mat_X_tmp}) has also been developped to improve the accuracy for highly connected graphs. The latter exploits the fact that in modern linear algebra libraries like Numpy the matrix multiplication is faster as the matrices are sparser. When the graphs are too dense it is thus interesting to use their complements, and we define the \textit{complement incidence matrix} $\bar{R}_\ast$ as the unoriented incidence matrix of the complement graph $\bar{G}_\ast$. However, the complements cannot be used for one graph and not the other, so the switching criterion has to be globally defined based on the densities of both graphs. Let the \textit{incidence matrix} $\tilde{R}_\ast$ be:
\begin{align}
\tilde{R}_\ast & = \begin{cases}
  R_\ast, & \text{if}\ 4(m_A + m_B) \leq n_A(n_A+1) + n_B(n_B+1) \\
  \bar{R}_\ast, & \text{otherwise}
\end{cases}
\label{eq:def_incidence}
\end{align}

The update equations can then be rewritten:
\begin{align}
Y_{k} & = \tilde{R}_A^\top X_{k-1} \tilde{R}_B \label{eq:ug_update_mat_Y} \\
X_{k} & = \tilde{R}_A Y_k \tilde{R}_B^\top \label{eq:ug_update_mat_X}
\end{align}

Note that edge attributes cannot be preserved with graph complements, which might look like a severe incompatibility with the present algorithm as it is precisely designed to account for all graph attributes. However, only the initialization equation~(\ref{eq:ug_init_mat}) uses the edge distance matrix $E$, which contains all the information about the similarities of edges attributes. Interestingly, as long as this information is injected during the initialization step it propagates as well in the complements. Using complements in the GPU implementation would requires to transfer more graphs and the speed gain is not obvious in this case, so the switch to complements has been only implemented on the CPU version.

\subsection{Iterative procedure for directed graphs}
\label{sec:iter_dg}

Let us now cover the case where both $G_A$ and $G_B$ are directed. For any edge $i$, let $s(i)$ and $t(i)$ be its source and target vertices, respectively. As previously, at iteration $k$ the vertex similarity score is $x_{uv}(k)$  and the edge similarity score is $y_{ij}(k)$. For the initial step, the vertex scores are defined as:

\begin{equation}
x_{uv}(1) = \left( \nu_{uv} + h_{uv}\right) \left( \sum_{\substack{s(i)=u \\ s(j)=v}} \epsilon_{ij} + \sum_{\substack{t(i)=u \\ t(j)=v}} \epsilon_{ij} \right)
\label{eq:dg_init_dvlp}
\end{equation}
 
And the update equations are, for $k>1$:

\begin{align}
y_{ij}(k) & = \frac{1}{f_y} \left( x_{s(i)s(j)}(k-1) + x_{t(i)t(j)}(k-1)\right)
\label{eq:dg_update_dvlp_Y}\\
x_{uv}(k) & = \frac{1}{f_x} \left( \sum_{\substack{s(i)=u \\ s(j)=v}} y_{ij}(k) + \sum_{\substack{t(i)=u \\ t(j)=v}} y_{ij}(k) \right)
\label{eq:dg_update_dvlp_X}
\end{align}

\noindent where $f_x$ and $f_y$ are the normalization coefficients, set at $1$ for further equations as previously. Note that in eq.(\ref{eq:dg_update_dvlp_X}) there is a difference as compared to the directed case and eq.(\ref{eq:ug_update_dvlp_X}) in the handling of self-loops: they are counted twice -- once as a source and once as a target -- while they are counted only once in the undirected case.

As introduced in \cite{Zager_2008}, it is  convenient to represent the adjacency structure of the graphs by pairs of matrices termed the \textit{source-edge matrix} $S_\ast$ and \textit{terminus-edge matrix} $T_\ast$, which are akin to the incidence matrix $R_\ast$ in the undirected case and defined as follows:
\begin{align}
[S_\ast]_{ui} & = \begin{cases}
  1, & \text{if}\ s(i)=u \\
  0, & \text{otherwise}
\end{cases}
\\
[T_\ast]_{ui} & = \begin{cases}
  1, & \text{if}\ t(i)=u \\
  0, & \text{otherwise}
\end{cases}
\label{eq:def_d_ST}
\end{align}

The adjacency matrix can then be recovered with $\Lambda_\ast = S_\ast T_\ast^\top$, and the incidence matrix of the corresponding undirected graph is simply $R_\ast = S_\ast \lor T_\ast$. As for undirected graphs, the graph complements may be used for the update equations and we define:
\begin{align}
(\tilde{S}_\ast, \tilde{T}_\ast)  & = \begin{cases}
  (S_\ast, T_\ast), & \text{if}\ 2(m_A + m_B) \leq n_A^2 + n_B^2 \\
  (\bar{S}_\ast, \bar{T}_\ast), & \text{otherwise}
\end{cases}
\label{eq:def_ST}
\end{align}

The scores initialization and update equations are then:
\begin{align}
X_{1} & = (V + H) \odot (S_A E S_B^\top + T_A E T_B^\top) \label{eq:dg_init_mat} \\[10pt]
Y_{k} & = \tilde{S}_A^\top X_{k-1} \tilde{S}_B + \tilde{T}_A^\top X_{k-1} \tilde{T}_B \label{eq:dg_update_mat_Y} \\
X_{k} & = \tilde{S}_A Y_k \tilde{S}_B^\top + \tilde{T}_A Y_k \tilde{T}_B^\top \label{eq:dg_update_mat_X}
\end{align}

\subsection{Convergence criterion}

Convergence has been extensively discussed in the works that inspired the present algorithm, and the demonstration of convergence for GASM is the same as for Zager's algorithm \cite{Zager_2008}. However, an important aspect that has not been investigated so far is the number of iterations before convergence. Since for GASM the iteration time strongly depends on the number of edges in the graphs, avoiding unnecessary iterations is critical for large and dense graphs.


Let us propose an estimated convergence criterion based on \textit{ad hoc} properties of the graphs. Considering that each iteration propagates the structural and attributes information one vertex further, the minimal number of iterations before every pair of vertices receives some information from all the other pairs of vertices is:

\begin{equation}
\tilde{k} = min(\Delta_A, \Delta_B)
\label{eq:convergence}
\end{equation}

\noindent where $\Delta_\ast$ is the diameter of $G_\ast$, \ie the maximum eccentricity among all vertices. For directed graphs, a safer definition is to use the diameter of the undirected versions of the graphs, but it seems that this is not necessary in practice so this is not what is used here.

Interestingly, as the graph diameter of fully connected graphs decreases when the density of edges increases (Supp. fig.~\ref{supp:k_star}), the dense graphs -- which have a higher iteration time -- benefit from a faster convergence and a reduced number of iterations. Convergence is achieved in just a few iterations with small-world graphs: the isomorphic matching of an Erdös-Rényi (ER) $G_{np}$ graph with 100 vertices and no attributes requires less than $\tilde{k}=4$ iterations, and it is reduced down to $\tilde{k}=2$ when the average degree is above 3 (Supp. fig.~\ref{supp:convergence_self_ER}).

\subsection{Matching}

The final step of scores determination is to handle isolated vertices, \ie vertices disconnected from the rest of the graph. These vertices may have attributes to be matched on, but at each iteration their scores in $X_k$ are all set to zeros. So, in order to take them into acount in the matching, we restore their scores to their initial values in $V$ divided by the appropriate normalisation factor:

\begin{equation}
   \forall u, v: x_{u,v}(\tilde{k}) \leftarrow \nu_{u,v}/f_x^{\tilde{k}-1} \quad \text{ if } u \text{ is isolated or } v \text{ is isolated}
  \label{eq:isolated_vertices}
\end{equation}

Finally, the matching is performed on the vertex scores matrix $X_{\tilde{k}}$ with a standard LAP algorithm searching for a maximal global score. It can also be performed on the edge score matrix $Y_{\tilde{k}}$, in case it is more relevant.


\section{Results}
\label{sec:results}

\subsection{Handling indeterminacies}
\label{sec:degeneracy_handling}

The GASM algorithm has been greatly inspired by the work of Zager and Verghese \cite{Zager_2008}, but there are several differences that make it better at handling indeterminacies -- \ie situations where multiple matching solutions are possible. In the next paragraphs we distinguish three types of indeterminacies and explain how GASM optimizes the three cases as compared as the original algorithm.

\subsubsection{Indeterminacies due to local symmetries}
\label{sec:degeneracy_symmetries}

\begin{figure}[!t]
  \centering
  \includegraphics[width=1\columnwidth]{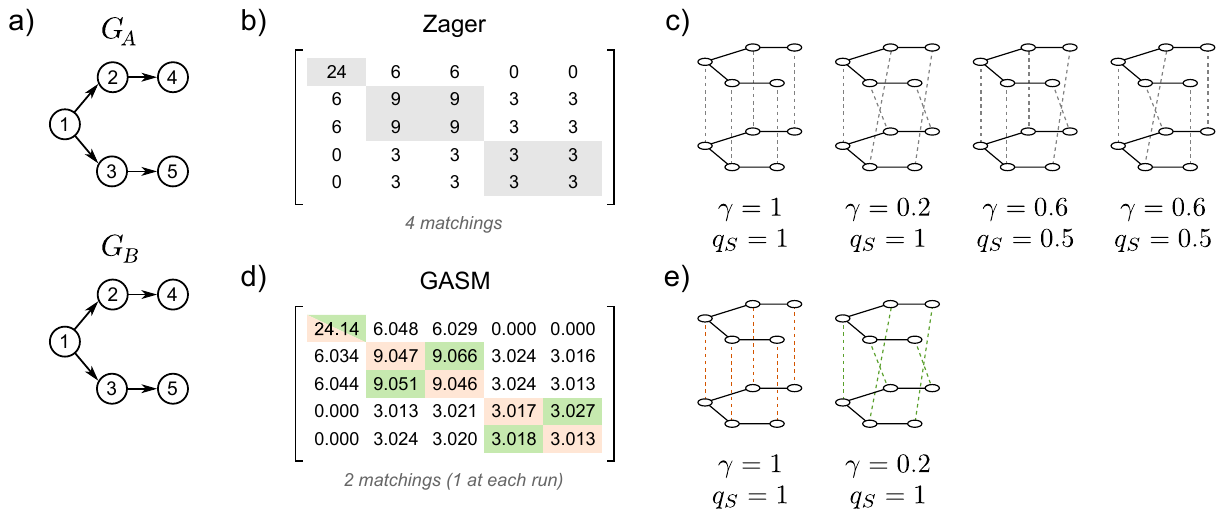}
  \caption{Managing symmetry with a minute noise.
  \textbf{a)} Self-matching ($G_A = G_B$) of a simple directed graph with a symmetry.
  \textbf{b)} Score matrix $X_{\tilde{k}=2}$ given by the algorithm of Zager \etal~\cite{Zager_2008}, without normalization. The matching solutions comprise only the grayed cells.
  \textbf{c)} The 4 corresponding matchings with the best score ($48$), along with their accuracy $\gamma$ and structural quality $q_S$.
  \textbf{d)} Exemple of score matrix $X_{\tilde{k}=2}$ produced by the GASM algorithm, without normalization and with a relatively large noise $\eta=10^{-2}$ to ease visualization. Here the best matching solution lies in the green cells, but other initial random numbers could favor the orange cells.
  \textbf{e)} The 2 corresponding matchings solutions.
  }
  \label{fig:ex_undet_1}
\end{figure}

Many graphs have local symmetries, \ie similar subgraphs attached to the rest of the graph with the same anchoring points. Local symmetries can take the form of branches, cycles, or more complex patterns. The indetermination comes from the fact that vertices at the same relative position in the symmetric sub-patterns have exactly the same surrounding structure, so with a structure-based scoring like Zager's their pairs have the same scores and the LAP solver has no way to determine which vertex belongs to which sub-pattern. 

One way to circumvent the problem could be to compute all the possible solutions of the LAP, and rank them by their structural quality for instance. Finding all the solutions of a LAP is P-complete \cite{Valiant_1979} and some algorithms are available for this task \cite{Fukuda1992,Uno1997} but they represent a consequent computational overhead and would make the matching of large ($n_\ast>100$) graphs intractable in practice. 

We have thus choosen a different approach, explained here with an exemple for the sake of clarity. Figure \ref{fig:ex_undet_1}-a depicts a basic branched graph for which self-matching with the Zager algorithm produces several pairs of vertices with similar scores  (fig.\ref{fig:ex_undet_1}-b) and 4 solutions are possible, 2 being structurally unsound. This situation is akin to the one previously reported in section~\ref{sec:matching quality}.

Interestingly, the addition of a minute random noise to the initial scores -- lying in the $h_{uv}$ term in equations (\ref{eq:ug_init_dvlp}) and (\ref{eq:dg_init_dvlp}) -- allows to filter out the solutions with low structural quality (fig.\ref{fig:ex_undet_1}-d,e). It may seem counterintuitive that adding some noise to the inputs can actually improve the outcome of a deterministic algorithm, so let us clarify how this works: first, the noise used in practice is so small ($\eta = 10^{-10}$) that it does not mess up with the general score determination and, for graphs without attribute, the integer part of the GASM score matrices is similar to the one obtained with Zager's aglorithm. This is also the case in fig.\ref{fig:ex_undet_1}-b,d even though the noise is much larger to ease visualization. Second, the noise cannot improve the average accuracy of the matchings: in the exemple of fig.\ref{fig:ex_undet_1}, the average accuracy of the solutions is $\gamma=0.6$ for both algorithms. However, it allows to filter out the solutions with low structural quality wherever there is a local symmetry. The mechanism at play is the following: initially the noise favors at random some pairs of vertices among the pool that would be otherwise degenerated. After a number of iterations corresponding to the symmetric sub-pattern size, the initial conditions have propagated and it is now pairs of whole sub-patterns that are slightly favored. After convergence, all the degeneracies have been raised and there is only one solution given by the score matrix. However, different solutions can still emerge from run to run, depending on the initial noise.

\subsubsection{Indeterminacies raised by propagating attribute information}
\label{sec:degeneracy_information}

\begin{figure}[!h]
  \centering
  \includegraphics[width=0.7\columnwidth]{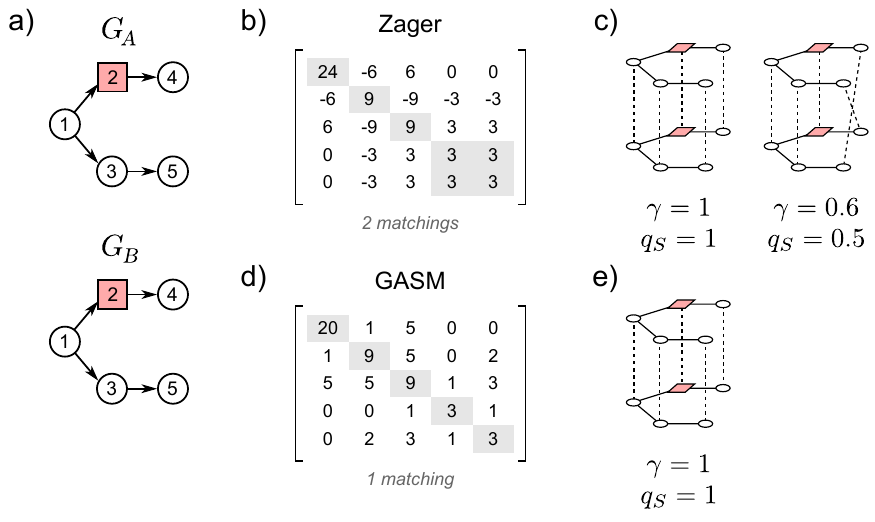}
  \caption{Propagation of attribute information through branches.
  \textbf{a)} Self-matching ($G_A = G_B$) of a simple directed branched graph with a categorical attribute on vertices. One vertex has a different value than the others, symbolized by a red square.
  \textbf{b)} Score matrix $X_{\tilde{k}=2}$ given by the Zager algorithm, without normalization. The matching solutions comprise only the grayed cells.
  \textbf{c)} The 2 corresponding matchings solutions with the best score ($48$), along with their accuracy $\gamma$ and structural quality $q_S$.
  \textbf{d)} Integer part of the score matrix $X_{\tilde{k}=2}$ produced by the GASM algorithm, without normalization. The decimal part, due to the artificial noise, is neglectible for the matching and is skipped to ease visualization.
  \textbf{e)} The corresponding matching solution.
  }  
  \label{fig:ex_undet_2}
\end{figure}

Attributes provide some information that is interesting to exploit. For categorical attributes, Zager \etal~proposed to multiply term-wised the converged, purely structural vertex score matrix with a distance matrix made of -1 and +1 to adjust the score matrix, as shown in the exemple of fig.\ref{fig:ex_undet_2}-b. Though this indeed raises the degeneracy for the concerned vertices, this approach suffers from the fact that neighboring vertices do not benefit from this information and can still be mismatched, as exemplified by the second solution in fig.\ref{fig:ex_undet_2}-c.

GASM introduces the attribute information in the initial score matrix $X_1$ \textit{via} the $V$ and $E$ matrices in eq.(\ref{eq:ug_init_mat}) and (\ref{eq:dg_init_mat}), so before the iterative procedure. This creates a coupling between the structure and the attributes during iterations which let the scores be determined not only by the similarity of the local structure and the vertices/edges proper attributes, but also by the similarity of the attributes of neighboring vertices and edges. In the exemple of fig.\ref{fig:ex_undet_2}, the solution where vertices 4 and 5 are mixed up is filtered out by GASM, which increases both the average accuracy and structural quality of the solutions. The mechanism of information propagation is actually similar to what has been described with noise in section \ref{sec:degeneracy_symmetries}, except that the initial differences are based on the attributes and thus do not change from run to run.

\subsubsection{Indeterminacies due to intrinsic differences in attributes}
\label{sec:degeneracy_uncompatible}

\begin{figure}[!h]
  \centering
  \includegraphics[width=1\columnwidth]{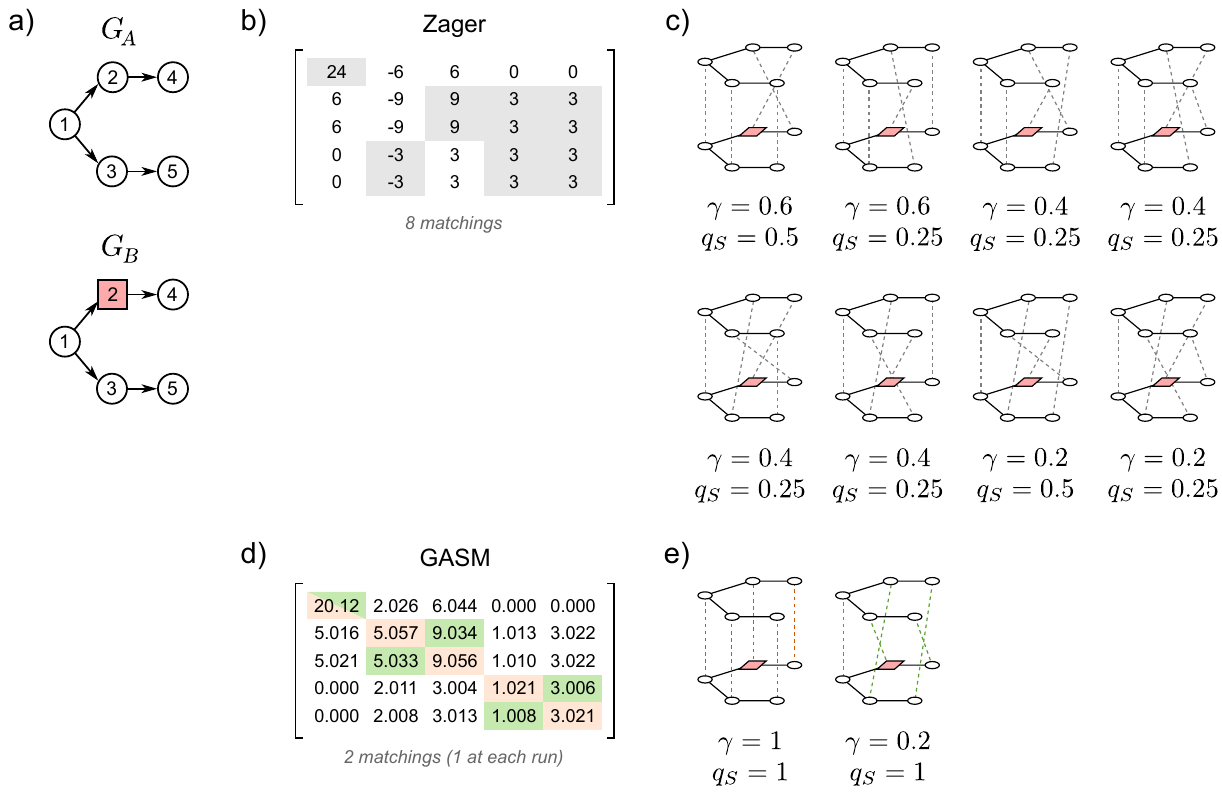}
  \caption{Managing intrinsic indeterminacies over attributes.
  \textbf{a)} The two graphs share the same structure but one vertex categorical attribute differs, symbolized by a red square.
  \textbf{b)} Score matrix $X_{\tilde{k}=2}$ given by the Zager algorithm, without normalization. The matching solutions comprise only the grayed cells.
  \textbf{c)} The 8 corresponding matchings solutions with the best score ($36$), along with their accuracy $\gamma$ and structural quality $q_S$.
  \textbf{d)} Exemple of score matrix $X_{\tilde{k}=2}$ produced by the GASM algorithm, without normalization and with a relatively large noise $\eta=10^{-2}$ to ease visualization. Here the best matching solution lies in the orange cells, but other initial random numbers could favor the green cells.
  \textbf{e)} The 2 corresponding matchings solutions.
  }
  \label{fig:ex_undet_3}
\end{figure}

Finally, outside the case of isomorphic matching the intrinsic differences between $G_A$ and $G_B$ can create several kinds of indeterminacies. Different structures are well-managed by both Zager's algorithm and GASM, but different attributes can make Zager's algorithm go totally wrong, as exemplified in fig.\ref{fig:ex_undet_3}: the vertices labelled 2 have different categories in $G_A$ and $G_B$ and the ($3\rightarrow5$) branch of $G_B$ could be equally matched with the ($2\rightarrow4$) and ($3\rightarrow5$) branches of $G_A$. In this exemple, the Zager algorithm leads to a large set of equally scored solutions (fig.\ref{fig:ex_undet_3}-c) which all seem unacceptable: the vertex 2 of $G_B$ is always matched with either the vertex 4 or the vertex 5 in $G_A$, which goes against the structural similarity of the graphs.

As show in fig.\ref{fig:ex_undet_3}-d,e, GASM finds all and only the structurally sound solutions. Attribute inconsistencies modify the initial score matrix $X_1$ by lowering the scores of the corresponding pairs in a symmetric way that do not affect the building up of scores based on the structural information. Indeterminacies leave a trace in the final score matrix -- see for instance the differences in the integer parts of the scores in fig.\ref{fig:ex_undet_1}-d and fig.\ref{fig:ex_undet_3}-d -- but it does not affect the emergence of the solutions based on structural cues. The noise plays the same determinant role as previously to filter out the solutions with low structural quality.

\subsection{Isomorphic matching}
\label{sec:isomorphic_matching}

\begin{figure}[!h]
  \centering
  \includegraphics[width=1\columnwidth]{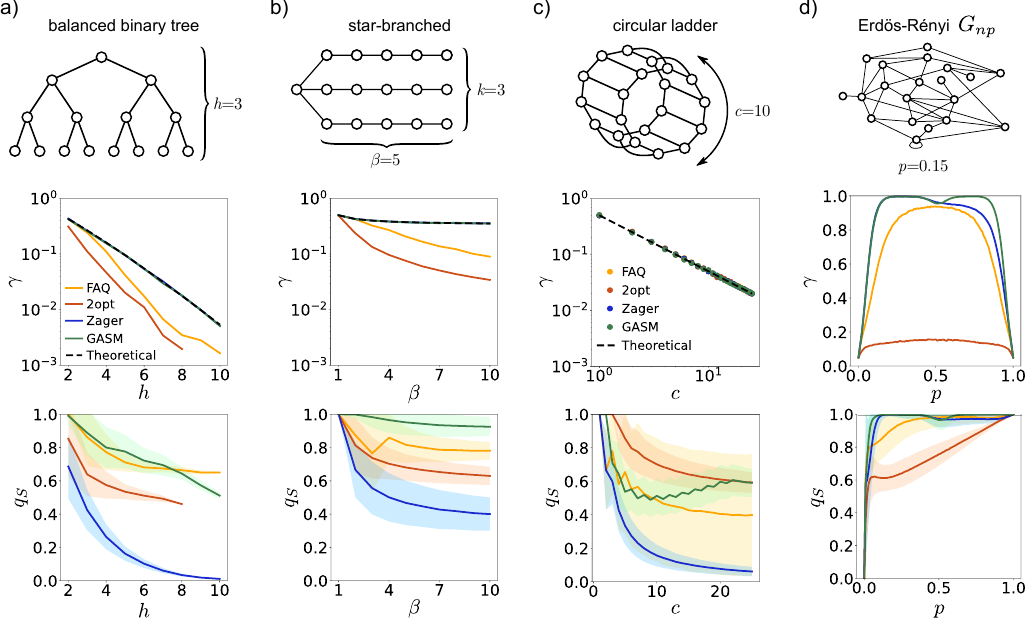}
  \caption{Isomorphic matching of different types of graphs: \textbf{a)} balanced binary trees with depth $h$, \textbf{b)}  star-branched with $k=3$ branches of length $\beta$, \textbf{c)} circular ladder with $2c$ vertices, and \textbf{d)} random Erdös-Rényi (ER) $G_{np}$ graphs with $n_A=20$ vertices and edge probability $p$. \textbf{Top}: Examples of each graph type. \textbf{Middle}: average accuracy $\gamma$ as a function of the graph parameters. \textbf{Bottom}: average structural quality $q_s$ computed over the same graphs. Colors are consistant in all panels. Each data point is averaged over $10^4$ samples, except for the balanced binary tree where it is variable with $h$ in order to keep a reasonnable computation time ; the data points for the 2opt algorithm are missing when $h>8$ due to a prohibitive computation time.
  }
  \label{fig:isomorphic_matching}
\end{figure}

Let us now delve deeper into the comparison of GASM with the other algorithms and the quantification of performance. For this we benchmarked 2opt, FAQ, Zager and GASM on both the average accuracy and structural quality over the same sets of graphs. Let us start with isomorphic matching, \ie the matching of two isomorphic graphs; in practice, $\Lambda_B$ is a shuffled version of $\Lambda_A$. The results for 4 types of undirected graphs are compiled in fig.\ref{fig:isomorphic_matching}.

Balanced binary trees (fig.\ref{fig:isomorphic_matching}-a) are a good exemple of graphs having multiple local symmetries and for which the maximal possible average accuracy $\gamma_{BT}$ can be determined analytically. Indeed, if there are $r$ vertices on a row they all have a $1/r$ probability to be correctly assigned, so there is on average one vertex correctly assigned per row and if $h$ is the depth of the binary tree there are on average $h+1$ vertices correctly matched in total. As the total number of vertices is $2^{h+1}-1$, the maximal possible average accuracy reads:

\begin{equation}
  \gamma_{BT} = \frac{h+1}{2^{h+1}-1}
  \label{eq:gamma_th_BT}
\end{equation}

While 2opt and FAQ have a poor accuracy on such graphs (fig.~\ref{fig:isomorphic_matching}-a middle), Zager and GASM stick to the theoretical maximum curve. To compare their solutions, one has to look at the structural quality $q_S$ (fig.~\ref{fig:isomorphic_matching}-a bottom), for which GASM has consistently higher values.

Star-branched graphs (fig.\ref{fig:isomorphic_matching}-b) also have a structure that allows to determine easily the maximum possible accuracy $\gamma_{SB}$. Similarly, there is on average one vertex correctly assigned per row and if $\beta$ is the branch depth there are on average $\beta+1$ vertices correctly matched. If there are $k$ branches, the total number of vertices is $k\beta+1$ and the maximal average accuracy reads:

\begin{equation}
  \gamma_{SB} = \frac{\beta+1}{k\beta+1}
  \label{eq:gamma_th_SB}
\end{equation}

Again, the accuracy of 2opt and FAQ drop as branches grow, while Zager and GASM always sit on the theorical maximum. We verified that this is true for virtually any values of $k$, and not just for $k=3$ as displayed in fig.\ref{fig:isomorphic_matching}-b. The tie on accuracy is broken by looking at the structural quality, which GASM dominates in all the tested range of parameters.

Third, circular ladders (fig.\ref{fig:isomorphic_matching}-c) are a very special family of graphs with respect to graph matching. First, the maximal possible average accuracy $\gamma_{CL}$ can as well be determined analytically: since all vertices have exactly the same surrounding structure, any vertex can be matched with any vertex and the best possible average accuracy simply reads: 

\begin{equation}
  \gamma_{CL} = \frac{1}{2c}
  \label{eq:gamma_th_CL}
\end{equation}

\noindent where $c$ is the number of vertices in a ring. Then, the minimal accuracy -- corresponding to random pairings -- is also equal to $1/2c$. It is thus expected that any algorithm would give solutions with the exact same accuracy, and indeed all four algorithms gave accuracies lying on the $1/2c$ curve. Again, the structural quality is useful to rank them: 2opt is better, followed by GASM, FAQ and finally Zager. When $c$ is large ($c \geq 22$) GASM becomes as good as 2opt.

All these graphs have many symmetries by construction, which may explain why GASM is particularly efficient on these datasets. Let's now turn to the isomorphic matching of random Erdös-Rényi $G_{np}$ graphs, which have much less symmetries. For such graphs the maximal possible accuracy is difficult to derive analytically, but the general idea is that when $p$ is close to $0$ or $1$ there are a lot of undeterminacies, \ie many vertices have the same surrounding structure and can be mismatched, while for intermediate values of $p$ there are much less undeterminacies and the maximal accuracy is close to $1$. Computation reveals that the average accuracy of 2opt is very low, and that FAQ is dominated by Zager, which is in turn below GASM. Apparently Zager is working better on graphs with low $p$, and the better accuracy of GASM at high $p$ is due to the complementing procedure described in section~\ref{sec:iter_ug}, which is typically activated when $p \gtrsim 0.5$. Again, the structural quality of the solutions is mostly dominated by GASM.

Overall, for these 4 types of graphs GASM features an excellent performance by consistently providing the best accuracy, and displaying the best structural quality for all types but circular ladders, where it is second. In addition, when attributes are present and lift degeneracies GASM is able to exploit these information and break the theoretical limitations due to the structural local symmetries; in this case, provided there are a sufficient amount of attributes and/or errors are low enough, there is no upper bound and perfect accuracy can, in principle, be achieved for any graph structure.

\subsection{QAPLIB benchmark}
\label{sec:QAPLIB}

\begin{figure}[!t]
  \centering
  \includegraphics[width=0.4\columnwidth]{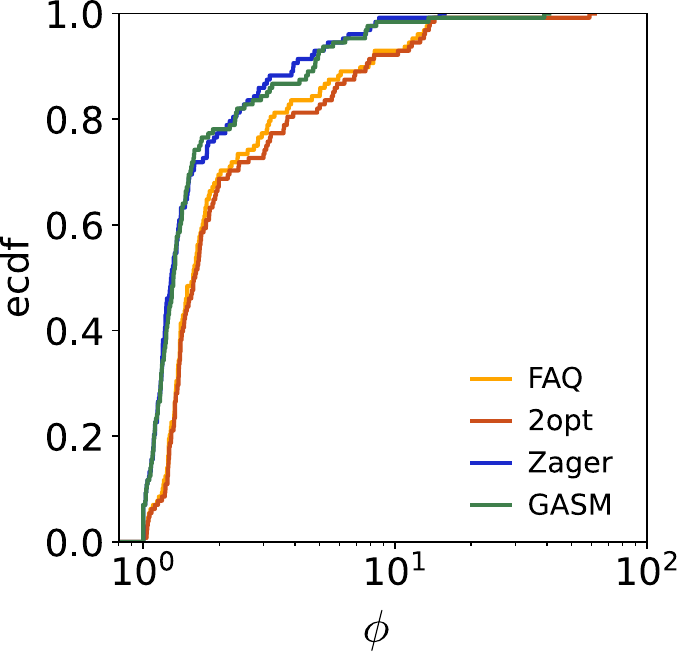}
  \caption{Benchmark on the QAPLIB database: empirical cumulative distributions over the QAPLIB instances of the score ratios $\phi$, obtained for different algorithms.}
  \label{fig:QAPLIB}
\end{figure}

Verifying the accuracy and structural quality for isomorphic matching is a good sanity check, but matching non-isomorphic graphs is a more realistic task. Let us simplify the problem by running benchmarks in two different contexts: the QAPLIB database, presented here, and graph degradation, which will be presented in the next section.

QAPLIB is a quadratic assignment problem library that has been widely used for benchmarking QAP algorithms \cite{Schellewald_2001, Zaslavskiy_2009}. It features 128 problems instances ranging in size from 10 to 256 vertices along with the best know solutions \cite{Burkard_1991}. Each problem comprises two matrices $A$ and $B$ (akin to $\Lambda_A$ and $\Lambda_B$ for graphs) and for each permutation $P$ a score can be computed as $tr(APB^\top P^\top)$. The QAP here is thus formulated to search for the \textit{minimal} score, while in the rest of this paper the similarity scores in $X$ and $Y$ were maximized. We computed the score ratios $\phi$ as the scores obtained for an algorithm divided by the score of the best known solution. One exception is the \verb|esc16f| instance, whose best known solution has a zero score. Since all algorithms found the minimal solution for this instance, the score ratios have been set to 1 for consistency.

The empirical cumulative distributions of $\phi$ for the 4 algorithms are show in fig.\ref{fig:QAPLIB}-a. In this representation the most leftwise curves have better solutions, and it is clear that Zager and GASM consistently found better solutions than 2opt and FAQ. As for the accuracy for isomorphic matching, the Zager and GASM algorithms provide very close scores. However, here the solutions cannot really be ranked in terms of structural quality due to the nature of the dataset: many of the instances correspond to fully-connected or other peculiar graphs, and the capacity of GASM to resolve local symmetries is largely irrelevant with QAPLIB.


\subsection{Graph degradation}
\label{sec:degradation}

Let us now turn onto a degradation benchmark, \ie a matching task where one graph is a degraded version of the other. There are actually many ways to degrade a graph: vertex swapping, edge swapping or flipping, adding noise to the attributes, \textit{etc.} but we cover here just the two major cases where \textit{i}) edges are removed or \textit{ii}) vertices are removed along with the corresponding edges, the latter task being also known as \textit{subgraph matching}. Note that the way elements are choosen for degradation is also determinant: it can be at random or in a given graph region, meaning that other regions are preserved. Here we will stick to the random case. In the next sections we assume without loss of generality that $G_B$ is a degraded version of $G_A$, with $n_A \ge n_B$ and $m_A \ge m_B$. The indices of the vertices of $G_B$ have also been systematically shuffled to avoid any accuracy bias.

\subsubsection{Edge removal}

Let us start with random edge removal, a task controlled by the degradation parameter $\delta_e$ defined as the amount of removed edges divided by the initial number of edges and that can be expressed as $\delta_e = 1-m_B/m_A$.

\begin{figure}[!ht]
  \centering
  \includegraphics[width=1\columnwidth]{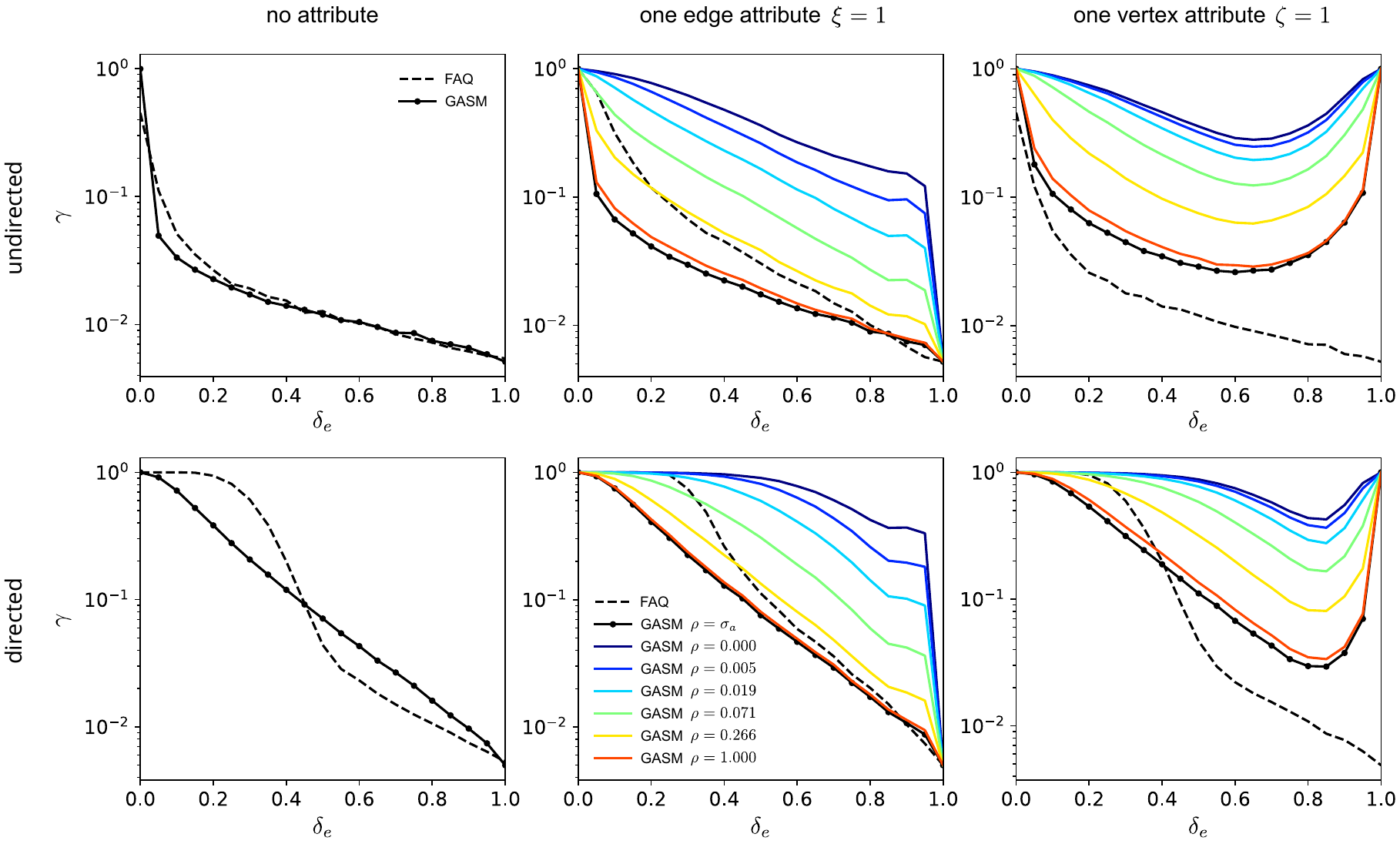}
  \caption{Edge degradation benchmark. Average accuracy $\gamma$ of FAQ and GASM plotted as a function of the edge degradation ratio $\delta_e$ for various conditions. Averaging is performed over $1,000$ ER $G_{np}$ graphs with $n=200$ vertices and $p_A=log(n)/n \simeq 0.01$. Top plots are for undirected graphs while bottom plots are for directed graphs. \textbf{Left}: graphs have no attribute. \textbf{Middle}: graphs have one measurable edge attribute ($\xi=1$) drawn from the standard normal distribution $\mathcal{N}(0,1)$. \textbf{Right}: graphs have one measurable vertex attribute ($\zeta=1$) drawn from the standard normal distribution. In this case, the FAQ curves correspond to no attribute and are displayed for reference. When an attribute is defined (middle, right), the default GASM attribute error (solid black) is the standard deviation of the difference of all attibrute pairs, $\rho=\sigma_a$ and colored curves correspond to manually defined attribute errors $\rho$.
  }
  \label{fig:degradation_edges}
\end{figure}

A comparison of the average accuracy of FAQ and GASM for the edge degradation of $G_{np}$ graphs is displayed in fig.\ref{fig:degradation_edges}, both for directed and undirected graphs. Without attribute, both algorithms see their accuracy drop steeply with $\delta_e$ for undirected graphs, while for directed graphs GASM has an exponential decay and FAQ features a very good tolerance to small degradations with an accuracy close to perfection up to $\delta_e=0.2$, though a rapid subsequent drop in accuracy place it below GASM for severe degradations ($\delta_e>0.5$).

The might of GASM becomes apparent when attributes can be exploited to improve the solutions. FAQ can also manage one measurable edge attribute, but do not take into account the error over this attribute. The middle panels of fig.\ref{fig:degradation_edges} compare FAQ and GASM accuracies in this situation ($\xi=1$), and the result highly depends on the error $\rho$ over the attribute, defined in (\ref{eq:m_attr_matrix}): high errors make GASM score poorly while low error turn it into an extremely degradation-tolerant algorithm. For instance the average accuracy for directed graphs with $\rho=0$ is still at $\gamma=0.9997$ when $\delta_e=0.5$, to be compared to $\gamma=0.1121$ for FAQ. This is not surprising since in that case GASM bases the matching primarily on the attribute information, and the structure is almost ignored.

A similar tendancy is obtained when there is one measurable vertex attribute ($\zeta=1$), as shown in the right panels of fig.\ref{fig:degradation_edges}. As FAQ cannot manage such an attribute, it is ignored and the average accuracy is the same as if there was no attribute. 

\subsubsection{Subgraph matching}

Let us now turn to vertex degradation, or subgraph matching. In this task, the graph $G_B$ is composed of a random subset of vertices of $G_A$, and only the edges of $G_A$ whose both vertices are in $G_B$ are kept. The subgraph task is thus parametrized by the properties of the inital graph $G_A$ and the degradation ratio $\delta_v$ defined as the amount of vertices removed divided by the initial number of vertices and that can be expressed as $\delta_v = 1-n_B/n_A$. The solution accuracy $\gamma$ is defined as the ratio of correctly matched vertices divided by the total number of vertices in the subgraph $n_B$.

\begin{figure}[!t]
  \centering
  \includegraphics[width=1\columnwidth]{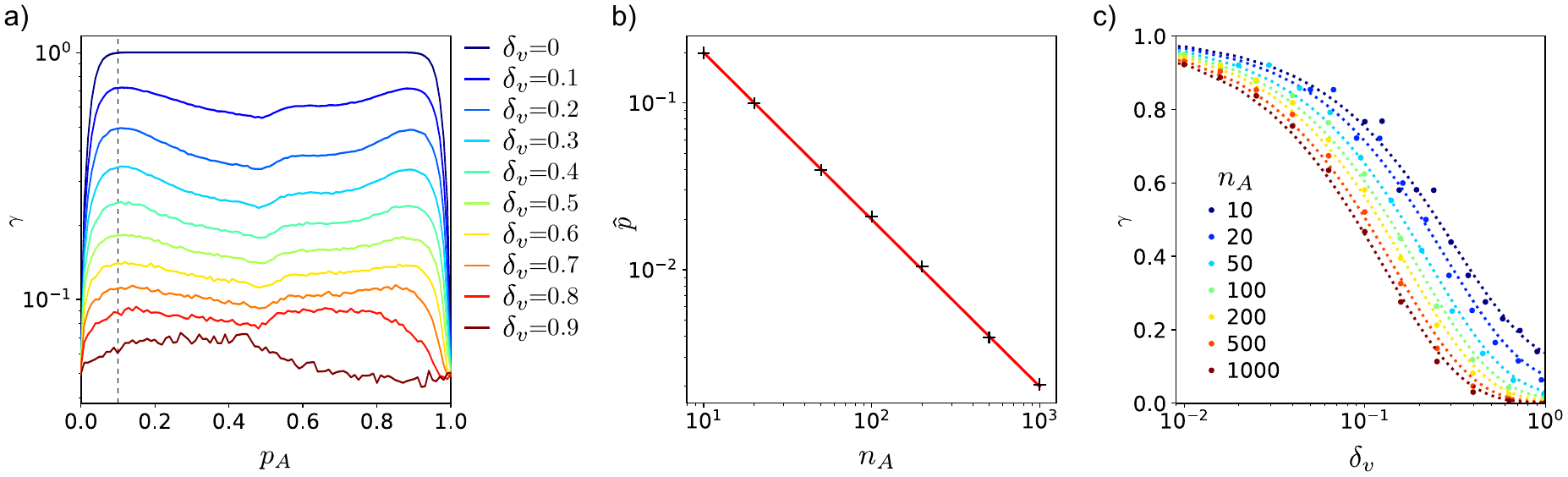}
  \caption{Accuracy peaks for directed ER graphs. \textbf{a)} Accuracy $\gamma$ as a function of the probability of edge creation $p_A$, for different vertex degradation ratios $\delta_v$ and $n_A=20$. All data points are averaged over $10^4$ realisations. Note that the curves with $\delta_v \in [0.1,0.8]$ all have a peak at the same value $\hat{p}$ (dashed line). A similar peak occurs at $1-\hat{p}$ because of the complement procedure described in section~\ref{sec:iter_ug} -- without it, the accuracy would monotonously decrease for $p_A>\hat{p}$.
  \textbf{b)} Scaling of $\hat{p}$ as a function of the initial number of vertices $n_A$ (black), fitted by $\hat{p}=2/n_A$ (red).
  \textbf{c)} Accuracy $\gamma$ as a function of the degradation ratio $\delta_v$ for different initial network sizes $n_A$ with $p_A=\hat{p}(n_A)$ (dots). Dotted curves correspond to fits given by eq.(\ref{eq:gamma_delta_v}).
  }
  \label{fig:p_star}
\end{figure}

For ER graphs, the initial graph $G_A$ is parametrized by the number of vertices $n_A$ and the edge ratio $p_A$. To reduce the number of parameters for the benchmark, we first tried to find the value of $p_A$ for which the accuracy is maximal. Indeed, Zager \etal~\cite{Zager_2008} observed that, for a few values of $p_A$, the accurary decayed when $p_A$ increased. However, at both limits $p_A=0$ (fully disconnected) and $p_A=1$ (fuly connected) no structural information can be infered and the accuracy has to drop to the minimal value of random matchings. So there has to be a maximum accuracy for some value of $p_A$ in $]0,1[$.

Figure~\ref{fig:p_star}-a shows the accuracy of GASM as a function of $p_A$ for different degradation ratios, and it appears that the peak location is essentially independant of $\delta_v$. Of course, when $\delta_v=0$ the subgraph $G_B$ is isomorphic to $G_A$ and the maximal possible accuracy is equal to one, except close to the extremal values of $p_A$ where the informational cut-off occurs. We then define $\hat{p}$ as the location of the accuracy peak in the range $\delta_v \in [0.1,0.8]$, and computed it for different sizes of the initial graph $n_A$ (fig.~\ref{fig:p_star}-b). The data points are nicely well-fitted by an inverse law, such that one can assess the empirical relationship:
\begin{equation}
  \hat{p} = \frac{2}{n_A}
  \label{eq:p_star}
\end{equation}

Since $p_A=d_A/n_A$, with $d_A=m_A/n_A$ being the average degree of $G_A$, this equation has a simple interpretation: the informational cut-off appears when $d_A \leq 2$, which corresponds to the threshold below which a graph is necessarily fragmented. Then, it is noticeable that the accuracy of GASM is well described by fits of the following form as a function of the degradation ratio $\delta_v$:

\begin{equation}
  \gamma(\delta_v) = \frac{1}{n_A} + \left( 1 - \frac{1}{n_A} \right) e^{-\delta_v/\alpha}
  \label{eq:gamma_delta_v}
\end{equation}

\noindent where $\alpha$ is a fit parameter (fig.\ref{fig:isomorphic_matching}-c).

\begin{figure}[!b]
  \centering
  \includegraphics[width=1\columnwidth]{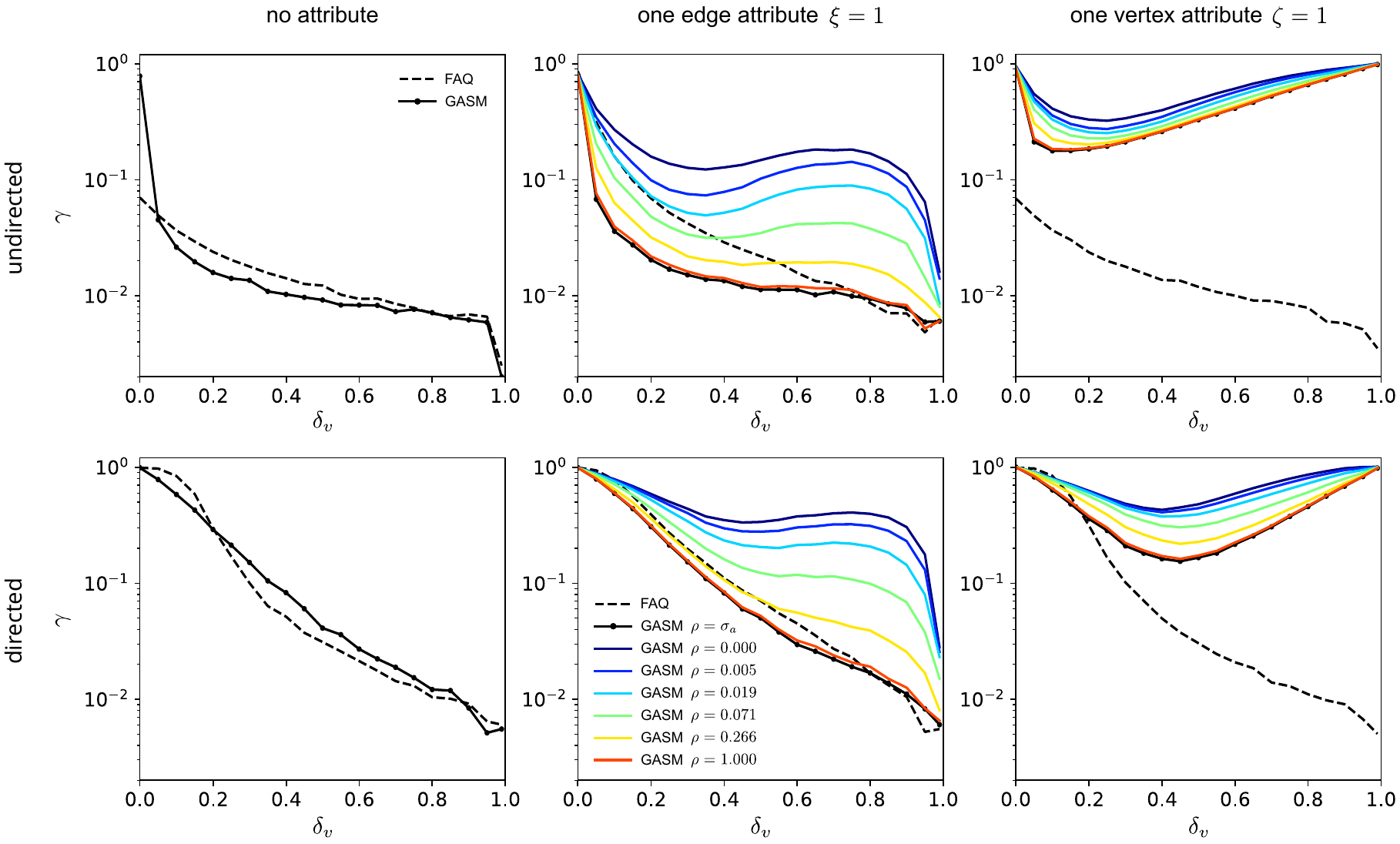}
  \caption{Vertex degradation benchmark. Average accuracy $\gamma$ of FAQ and GASM plotted as a function of the vertex degradation ratio $\delta_v$ for various conditions. Averaging is performed over $10^3$ ER $G_{np}$ graphs with $n_A=200$ vertices and $p_A=\hat{p}(n_A)=0.005$. Top plots are for undirected graphs while bottom plots are for directed graphs. \textbf{Left}: graphs have no attribute. \textbf{Middle}: graphs have one measurable edge attribute ($\xi=1$) drawn from the standard normal distribution $\mathcal{N}(0,1)$. \textbf{Right}: graphs have one measurable vertex attribute ($\zeta=1$) drawn from the standard normal distribution. In this case, the FAQ curves correspond to no attribute and are displayed for reference. When an attribute is defined (middle, right), the default GASM error (solid black) is the standard deviation of the difference of all attibrute pairs, $\rho=\sigma_a$ and colored curves correspond to manually defined errors $\rho$.
  }
  \label{fig:degradation_vertices}
\end{figure}

A comparison of the average accuracy of FAQ and GASM for the vertex degradation of ER $G_{np}$ graphs is displayed in fig.\ref{fig:degradation_vertices}, both for directed and undirected graphs. Without attribute, the general trend is very similar to what is observed for edge degradation in fig.\ref{fig:degradation_edges}-left, except for directed graphs where FAQ is less robust to small degradations and here the curves for both algorithm are much more similar.

Again, the ability of GASM to exploit attributes makes it more accurate than FAQ when the error is low enough, for both an edge attribute ($\xi=1$, fig.~\ref{fig:degradation_vertices}-middle) and a vertex attribute ($\zeta=1$, fig.~\ref{fig:degradation_vertices}-right). The dominance of GASM is particularly striking with one vertex attribute, since for all the tested errors the resulting accuracy is always at least one order of magnitude above FAQ.

\subsection{Speed}
\label{sec:speed}

\begin{figure}[!b]
  \centering
  \includegraphics[width=0.9\columnwidth]{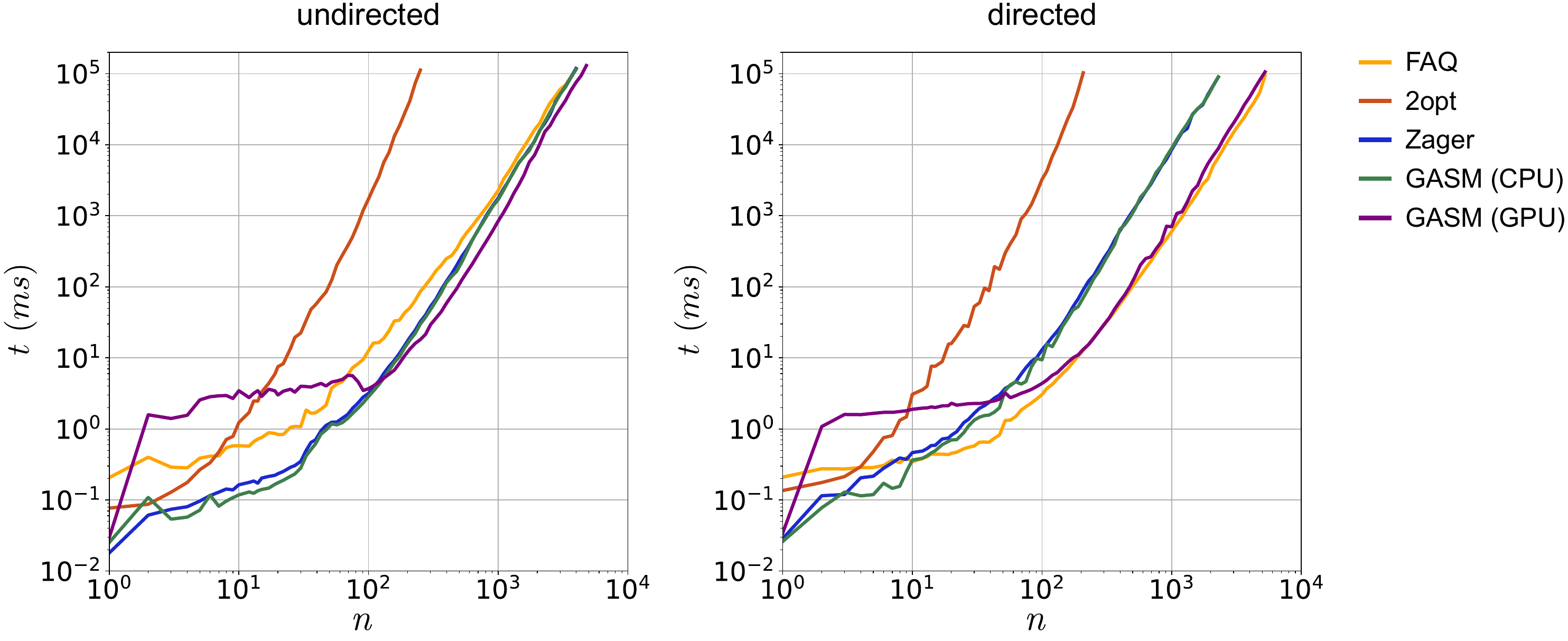}
  \caption{Timing benchmark on the isomorphic matching of ER $G_{np}$ graphs with $n$ vertices and $p=log(n)/n$. The benchmark has been performed on undirected (left) and directed (right) graphs, for 4 algorithms running on CPU and the implementation of GASM on GPU. All computing times are averaged over 10 realizations.}
  \label{fig:speed}
\end{figure}

As for many ressource-demanding problems, the search for approximate matching solutions faces a trade-off between accuracy and efficiency, the latter refering to computation time and memory resources. As observed in \cite{Vogelstein2015}, slow algorithms could probably achieve better accuracy given more time, and at the extreme an exhaustive search could reach optimal solutions at the cost of utmost time and memory budget. Put differently, accuracy and efficiency define a space where the best algorithms sit on a Pareto front a Pareto front and proper benchmarks should take both aspects into account.

However, computation speeds are difficult to compare for several reasons. Implementations and hardware constantly improve, and for instance there has been a speedup of 2 orders of magnitude from the first Matlab implementation of FAQ in 2015 \cite{Vogelstein2015} to our tests 10 years later with the current Scipy implementation running on a more recent computer. Even when one takes care to run a timing benchmark with algorithms written in the same language on the same computer, the high level of optimization of older algorithms makes the comparison with the first implementation of an emerging algorithm rather unfair. Finally, some algorithms are suitable for implementations on a GPU while others are not, and it is delicate to compare the timings when the hardware and technology are different.

With these limitations in mind, we conducted a timing benchmark for the isomorphic matching of ER $G_{np}$ graphs whose results are summarized in fig.~\ref{fig:speed}. Importantly, GASM is well-suited to GPU parallelization and, as explained in section~\ref{sec:iter_ug}, we implemented both a CPU version and a CUDA version. As with all GPU algorithms the data transfer between the host and the device has a cost, and for GASM it dominates the global running time for small graphs (typically below $10^2$ vertices). This is probably not an issue for many applications since for such small graphs the total running time remains below 10ms. 

The stakes are higher for larger graphs ($n>10^2$). In this range, the benchmark shows that: \textit{i}) 2opt is always slower by orders of magnitude, \textit{ii}) the CPU version of GASM is as fast as Zager in all cases, and \textit{iii}) slightly faster than FAQ for undirected graphs, but approximately 10 times slower for directed graphs. However, the GPU version of GASM is faster than FAQ for undirected graphs -- though the difference tends to vanish for large graphs -- and as fast as FAQ for directed graphs.

The previous sections indicated that GASM provides better solutions (in terms of accuracy, QAP scores and structural quality) than the other algorithms in most situations. The timing benchmark indicates that it is possible to have GASM running as fast or faster than the other algorithms, including FAQ, so altogether it seems to point out that GASM is more Pareto-optimal than all the other algorithms tested here.


\section{Conclusion}
\label{sec:conclusion}

This work presents the \textit{Graph Attributes and Structure Matching} (GASM) algorithm, which takes root in the iterative methods for graph matching. Notably, it improves the algorithm of Zager \etal~\cite{Zager_2008} in a number of ways:

\begin{itemize}
  \item it uses a minute noise to lift the degeneracies due to local symmetries,
  \item it implements a complement procedure, to take advantage of the fact that solutions are more accurate when the graphs are sparse,
  \item it handles properly isolated vertices,
  \item an \textit{ad hoc} convergence criterion is proposed,
  \item a GPU implementation has been implemented, which is particularly well-suited for this familly of algorithms,
  \item the integration of the attributes is done before the iterative procedure, which improves the quality of solutions and makes the algorithm more robust to discrepancies in the attributes,
  \item and most of all the ability to handle any number and types of attributes makes it well-suited to tackle real-world problems.
\end{itemize}

Importantly, GASM also introduces the notion of error on the attributes, which tunes \textit{in fine} how much the algorithm relies on the structure or on the attributes: if the attribute values are highly discriminant then GASM exploits principally this information, while attributes estimated with a large error only influence the solution search marginally.

Beyond the GASM algorithm, this study also formalizes the difference between categorical and measurable attributes and proposes a common framework to incorporate all these information. It also sheds light on the importance to systematically benchmark the undirected and directed cases separately, as we saw differences in all the measurements of our benchmarks, for all the tested algorithms. Moreover, we emphasize the importance of taking into account not only the accuracy /performance score of the solutions,  but also other measurements that are relevant in the context of graphs matching, like the structural quality for instance.

For future work, several leads can be explored. First, sometimes a partial matching of the vertices is known \textit{a priori}, and seeded graph matching has gained a lot of attention in the recent years \cite{Lyzinski_2014}. GASM can certainly be modified to leverage these information as well.

Then, there is room for further speed improvements. On the algorithmic side a better convergence criterion could considerably speed up the process by avoiding unnecessary iterations. It is clear from supp. fig.\ref{supp:convergence_self_ER} that there is still a lot of room for improvements on this matter; in addition, taking also into account the attributes for the convergence criterion should be also beneficial. Next, a fully on-GPU version is yet to implement, by porting on the device the computation of the initial score matrices, the complement procedure for dense graphs and finally the LAP solver itself. On the hardware side, the future use of PCIe 5.0 should in theory double the transfer rates and yet improve further the speed of GPU implementations.

Finally, the next scientific challenge is to apply GASM to real datasets, like for instance the comparison of protein-protein interaction networks, connectomes and artificial neural networks.




\bibliography{bibtex}
\bibliographystyle{abbrvurl}


\newpage
\begin{center}
\textbf{\Large Supplementary Materials}
\end{center}
\setcounter{equation}{0}
\setcounter{figure}{0}
\setcounter{table}{0}
\setcounter{page}{1}
\makeatletter
\renewcommand{\theequation}{S\arabic{equation}}
\renewcommand{\thefigure}{S\arabic{figure}}
\renewcommand{\thetable}{S\arabic{table}}
\setcounter{section}{0}
\renewcommand{\thesection}{S\arabic{section}}

\section{Computational details}

\subsection{Hardware and software}

All the code used for this article has been written in Python and is available in the following repository:

\url{https://github.com/CandelierLab/GraphMatching.git}

Scipy's optimized routines have been used as much as possible, notably the LAP solver in \verb+scipy.optimize.linear_sum_assignment()+, which implements the algorithm in~\cite{Crouse_2016}, and the QAP solver \verb+scipy.optimize.quadratic_assignment()+ which implements both the FAQ~\cite{Vogelstein2015} and 2opt~\cite{Fishkind_2019} algorithms. The Zager algorithm~\cite{Zager_2008} and GASM (CPU version) have been written to rely extensively on Numpy's optimization. The GPU version of GASM uses Numba to define the CUDA kernels. The timing benchmark was realized with the \verb+perf_counter_ns()+ function of Python's \verb+time+ module.

The timing benchmark have been performed on a single machine with the following specifications: Motherboard Asus ROG Maximus Z790 Formula, Intel Core i9-13900KS processor with 192Go of DDR5 RAM (5200 MHz, CL38) and a PNY Nvidia RTX A4500 graphics card.

\subsection{Approximate normalization factor.}
\label{sec:normalization}

Equations (\ref{eq:ug_update_dvlp_Y}-\ref{eq:ug_update_dvlp_X}) and (\ref{eq:dg_update_dvlp_Y}-\ref{eq:dg_update_dvlp_X}) define the normalization coefficients $f_x$ and $f_y$. From a formal point of view, these coefficients can be set to any strictly positive value at each step without altering the outcome of the algorithm, so they are removed in the subsequent equations for readability. 

However, in practice it can be dangerous to set $f_x = f_y = 1$ because the values of the score matrices increase exponentially with the iterations and may cause either a floating point overflow or precision issues related to the unit of least precision. For instance, when two graphs with an average degree of 500 are matched, the scores increase by a factor of the order of $10^6$ at each iteration. To reduce transfer and computation times the GPU version of GASM uses single-precision floats, which overflow at approximately $2^{128} \simeq 3.10^{38}$, so the overflow would occur in just 7 iterations. Also, the first integer that is not exactly representable is $2^{24}+1 \simeq 1.7 \times 10^7$, so precision issues may start within the first iterations.

One way to avoid these issues is to normalize the score matrices by their mean value (or any other norm) at each iteration as in \cite{Zager_2008}, but determining the mean scores is a computational overhead that can be avoided. First, only one normalization per iteration is enough and we can safely ignore the normalization of the edge scores $Y_k$, set $f_y=1$ and normalize only the scores in $X_k$. Then, only a rough estimate of the normalization coefficient is necessary for keeping the scores in a reasonnable range, and an \textit{ad hoc} estimation based on the graphs' average degree can be formulated as follows, both for directed and undirected graphs:

\begin{align}
  f_x = max(4 d_A d_B, 1)
  \label{eq:normalization_1}
\end{align}

\noindent where $d_\ast$ is the average degree of graph $G_\ast$ (\textit{outdegree} for directed graphs). It can then be slightly improved into the following form:

\begin{equation}
  f_x = 4 d_A d_B + 1
  \label{eq:normalization}
\end{equation}

\begin{figure}[!t]
  \centering
  \includegraphics[width=1\columnwidth]{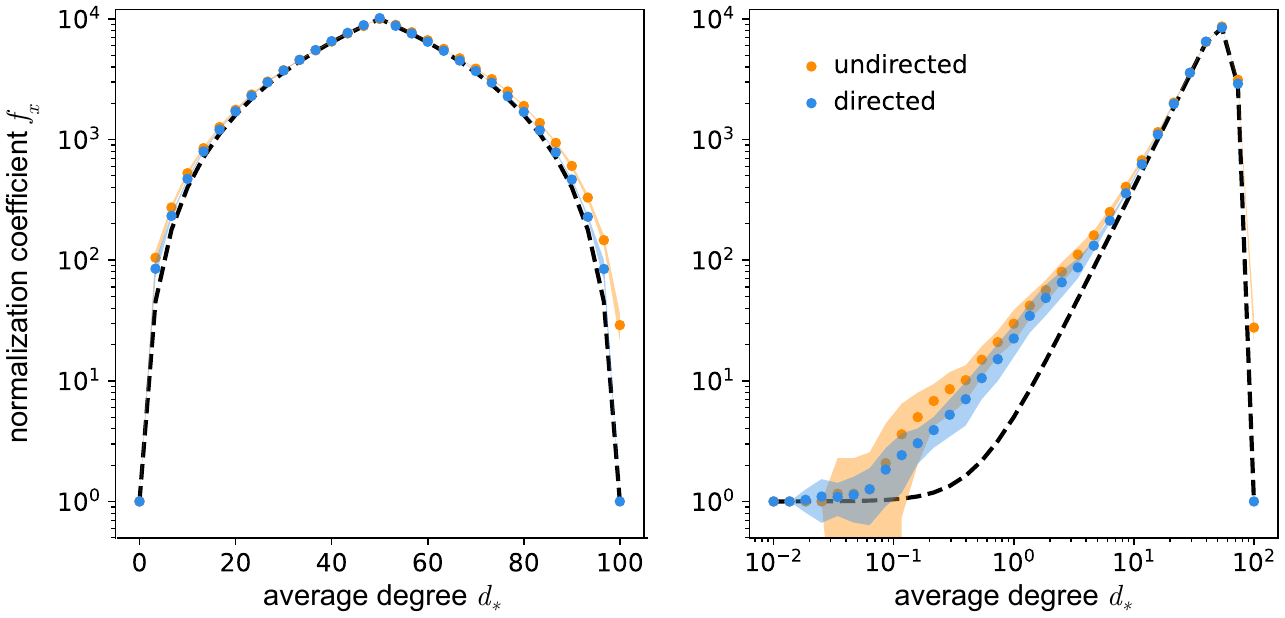}
  \caption{Estimated and approximated normalization coefficients $f_x$ for ER graphs with $n_\ast=100$ vertices in the isomorphic matching task, as a function of the average degree of the graphs $d_\ast$ with linear (left) and logarithmic (right) scales. The estimation of $f_x$ (dots) is defined as the ratio between the mean values of the vertex score matrix $X$ during the last 2 iterations before convergence, \ie $\left< X_{\tilde{k}} \right> / \left< X_{\tilde{k}-1} \right>$, averaged over 100 independant runs for directed (blue) and undirected graphs (orange). Shaded areas represent standard deviations. The dashed curve is the approximated normalization coefficient proposed in eq.~(\ref{eq:normalization}). The graph is symmetric due to the complement procedure described in section~\ref{sec:iter_ug}.}
  \label{supp:normalization}
\end{figure}

A comparison with estimated coefficients for ER $G_{np}$ graphs is provided in supp. fig.~\ref{supp:normalization}, showing that the error remains below a factor $10$. All the results presented in this article have been computed with the normalizaton coefficient provided by eq.~(\ref{eq:normalization}).

\section{Graphs}

\subsection{Random graphs}

The random graphs used in the paper are Erdös-Rényi-Gilbert $G_{np}$ graphs, which are constructed by defining a set of $n$ vertices and including every possible edge with probability $p$, independently from every other edge \cite{Fienberg_2012}.

\subsection{QAPLIB}

The instances and solutions of the QAPLIB library have been downloaded from \href{https://coral.ise.lehigh.edu}{https://coral.ise.lehigh.edu}.

\begin{figure}[!h]
  \centering
  \includegraphics[width=0.8\columnwidth]{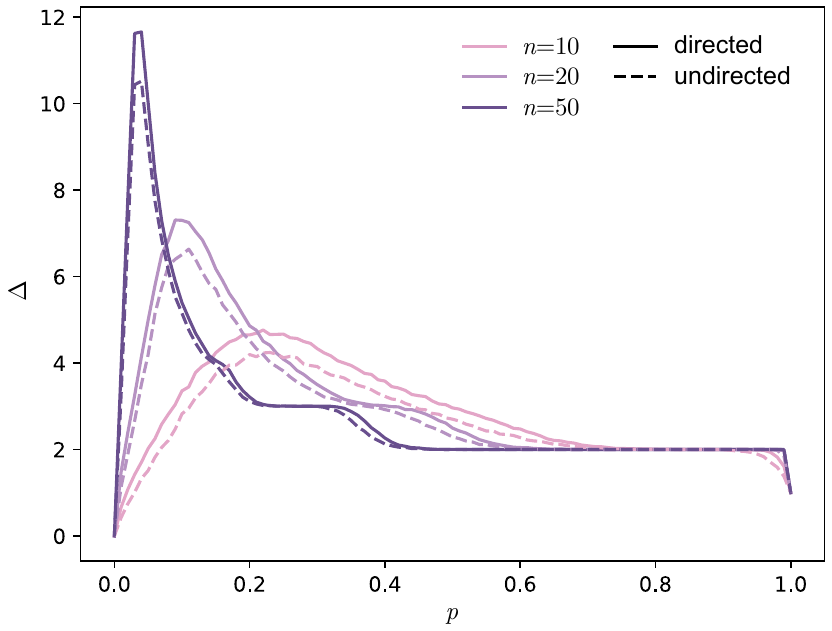}
  \caption{Average diameter $\Delta$ of directed and undirected Erdös-Rényi $G_{np}$ graphs as a function of $p$, for different values of $n$. Each point is averaged over 1,000 graphs.}
  \label{supp:k_star}
\end{figure}

\begin{figure}[!h]
  \centering
  \includegraphics[width=0.8\columnwidth]{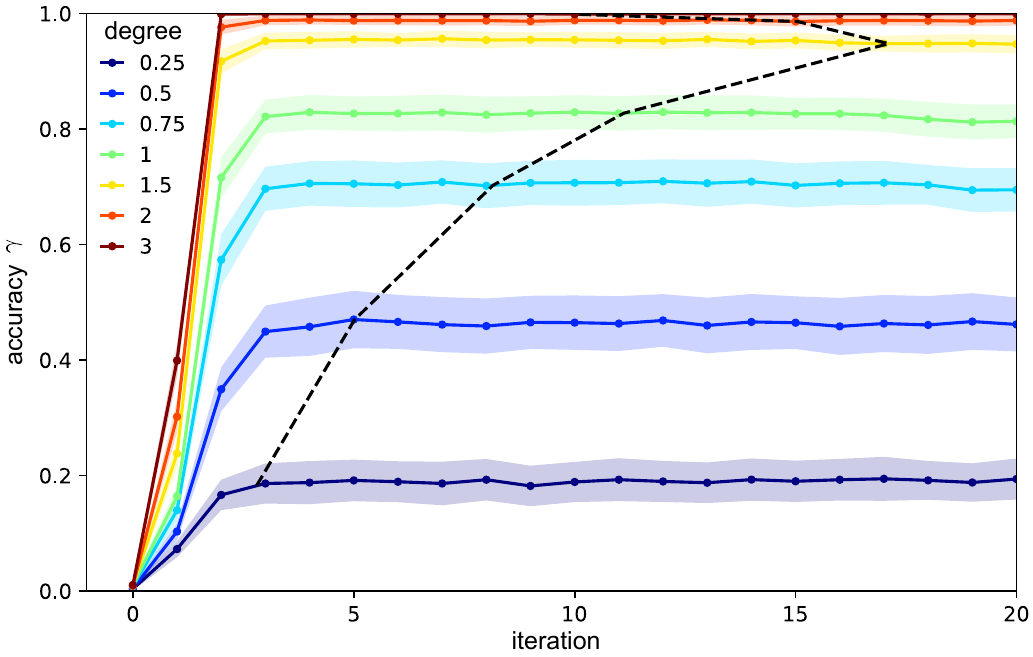}
  \caption{Convergence of GASM. Accuracy $\gamma$ during isomorphic matching of ER graphs with $n_\ast=100$ vertices and no attribute, for different average degree. Accuracies are averaged over $100$ runs, and the standard deviations are represented by the shaded areas. Iteration 0 represent uniform scores (random matching), and iteration $k \geqslant 1$ relies on the score matrix $X_k$. The dashed black curve indicates the average $\tilde{k}$ for the different degrees.}
  \label{supp:convergence_self_ER}
\end{figure}

\end{document}